\documentclass[pre,aps,twocolumn,showpacs,floatfix]{revtex4}
\usepackage[utf8]{inputenc}
\usepackage{cmap}
\usepackage[T1]{fontenc}

\usepackage{graphicx}
\usepackage{amsmath}
\usepackage{amsthm, amssymb, color}
\usepackage{sistyle}

\usepackage[acronym]{glossaries}
\newacronym{kpz}{KPZ}{Kardar--Parisi--Zhang}
\newacronym{ew}{EW}{Edwards--Wilkinson}
\newacronym{rsos}{RSOS}{restricted solid-on-solid}
\newacronym{bd}{BD}{ballistic deposition model}
\newacronym{ms}{MS}{multisurface coding}
\newacronym{fdr}{FDR}{fluctuation-dissipation relation}
\newacronym{kk}{KK}{Kim--Kosterlitz}
\newacronym{mcs}{MCS}{Monte-Carlo steps}

\newacronym{rng}{RNG}{random number generator}
\newacronym{gpu}{GPU}{graphics processing unit}
\newacronym{gpgpu}{GPGPU}{general purpose computing on graphics processing units}
\newacronym{cpu}{CPU}{central processing unit}
\newacronym{simt}{SIMT}{single instruction multiple thread}

\newacronym{dd}{DD}{domain decomposition}
\newacronym{cb}{CB}{checker-board}
\newacronym{sca}{SCA}{stochastic cellular automaton}
\newacronym{rs}{RS}{random-sequential}
\newacronym{dt}{DT}{double tiling}
\newacronym{db}{DB}{dead border}

\newacronym{pl}{PL}{power law}
\newacronym{rhs}{r.h.s.}{right-hand side}
\newacronym{resp}{resp.}{respectively}
\newacronym{snr}{S/N}{signal-to-noise ratio}
\newacronym{fom}{FOM}{figure of merit}
\makeglossaries

\newcommand\kpzRsosAlpha{\num{.390}(4)}

\begin{document}

\title{Universality of (2+1)-dimensional restricted solid-on-solid models}

\author{ Jeffrey Kelling (2,3), G\'eza \'Odor (1) and Sibylle Gemming (3,4)}

\affiliation{
(1) Institute of Technical Physics and Materials Science,
Centre for Energy Research of the Hungarian Academy of Sciences \\
P.O.Box 49, H-1525 Budapest, Hungary \\
(2) Department of Information Services and Computing, \\
Helmholtz-Zentrum Dresden-Rossendorf \\
P.O.Box 51 01 19, 01314 Dresden, Germany\\
(3) Institute of Ion Beam Physics and Materials Research \\
Helmholtz-Zentrum Dresden-Rossendorf \\
P.O.Box 51 01 19, 01314 Dresden, Germany\\
(4) Institute of Physics, TU Chemnitz\\
09107 Chemnitz, Germany}

\begin{abstract}
Extensive dynamical simulations of restricted solid on solid
models in $D=2+1$ dimensions have been done using parallel
multisurface algorithms implemented on graphics cards.
Numerical evidence is presented that these models
exhibit \glsdesc{kpz} surface growth scaling, irrespective
of the step heights $N$. We show that by increasing $N$ the
corrections to scaling increase, thus smaller step sized
models describe better the asymptotic, long-wave-scaling
behavior.
\end{abstract}
\maketitle

\section{Introduction}

The \gls{kpz} equation~\cite{PhysRevLett.56.889} describes the
evolution of a fundamental, non-equilibrium surface growth model
by a Langevin equation
\begin{equation}  \label{KPZ-e}
\partial_t h(\mathbf{x},t) = \sigma \nabla^2 h(\mathbf{x},t) +
\lambda(\nabla h(\mathbf{x},t))^2 + \eta(\mathbf{x},t) \ .
\end{equation}
The scalar field $h(\mathbf{x},t)$ is the height, progressing
in the $D$ dimensional space relative to its mean position, that
moves linearly with time $t$.
A smoothing surface tension is represented by the coefficient $\sigma$,
which competes a curvature-driven propagation, described by the nonlinear coefficient
$\lambda$ and a zero-average Gaussian stochastic noise.
This noise field exhibits the variance
$\langle\eta(\mathbf{x},t)\eta(\mathbf{x^{\prime}},t^{\prime})\rangle =
2 \Gamma \delta^D (\mathbf{x-x^{\prime}})(t-t^{\prime})$,
with an amplitude, related to the temperature in the equilibrium system,
and $\langle\rangle$ denotes a distribution average.
Besides describing the dynamics of simple growth processes \cite{H90}
\gls{kpz} was inspired in part by the stochastic Burgers equation
\cite{Burgers74} and is applicable for randomly stirred fluids \cite{forster77},
for directed polymers in random media \cite{kardar85} for
dissipative transport \cite{beijeren85,janssen86} and for the
magnetic flux lines in superconductors \cite{hwa92}.

Discretized versions have been studied frequently over the past few decades
\cite{meakin,barabasi,krug1997review}.
The morphology of a surface of linear size $L$ can be
described by the squared interface width
\begin{equation}
\label{Wdef}
W^2(L,t) = \frac{1}{L^2} \, \sum_{i,j}^L \,h^2_{i,j}(t)  -
\Bigl(\frac{1}{L^2} \, \sum_{i,j}^L \,h_{i,j}(t) \Bigr)^2 \ .
\end{equation}
In the absence of any characteristic length simple
growth processes are expected to be scale-invariant
\begin{equation}
\label{FV-forf}
W(L,t) \propto L^{\alpha} f(t / L^z),
\end{equation}
with the universal scaling function $f(u)$
\begin{equation}
\label{FV-fu}
f(u)  \propto
\left\{ \begin{array}{lcl}
     u^{\beta}     & {\rm if} & u \ll 1 \\
     {\rm const.} & {\rm if} & u \gg 1
\end{array}
\right .
\end{equation}
Here $\alpha$ is the roughness exponent in the stationary regime,
when the correlation length has exceeded $L$ and
$\beta$ is the growth exponent, describing the intermediate time
behavior. The dynamical exponent $z$ can be expressed as the ratio
of the growth exponents
\begin{equation}\label{zlaw}
z = \alpha/\beta \
\end{equation}
and due to the Galilean invariance the $\alpha+z=2$ relation
holds as well.

While in $D=1+1$ exact solutions are known, due to the Galilean symmetry
\cite{forster77} and an incidental fluctuation-dissipation symmetry
\cite{kardar87}, in higher dimensions \gls{kpz} has been investigated by
various analytical \cite{KCW,SE92,FT94,L95,F05,CCDW11} and numerical methods
\cite{FT90,PhysRevLett.109.170602,MPP2000,FDAAR05}, still debated issues remain.
For example, there is a controversy on
the surface growth exponents of the $D=2+1$ \gls{kpz}, obtained by recent
simulations \cite{PhysRevE.84.061150,AOF2014} and a field theoretical
study \cite{L98}.
Assuming that the height correlations do not exhibit multi-scaling
and satisfy an operator product expansion Ref.~\cite{L98} concluded
that growth exponents are rational numbers in two and three
dimensions \cite{L98}. This was in accordance with some earlier
\gls{rsos} model simulation results \cite{kimKosterlitz1989,Kim91}.
Recent high precision simulations 
\cite{Halp13,AOF2014_critDim,PhysRevE.84.061150,AOF2014,PaganiParisi2015}
all excluded this and concluded
$\alpha=0.393(4)$~\cite{PhysRevE.84.061150,PaganiParisi2015,AOF2014} and
$\beta=0.2414(15)$~\cite{PhysRevE.84.061150}.
\gls{rsos} models are defined by deposition at random sites if the
local height difference satisfies
\begin{equation}
\vert  h(\mathbf{x},t) -  h(\mathbf{x'},t)\vert \le N \ .
\end{equation}

Very recently Kim \cite{kim2015} investigated \gls{rsos} models with maximum step sizes
$N=1,2,\ldots,7$. As he increased $N$ the roughness exponent $\alpha$ seemed
to converge to $4/10$ and the growth exponent $\beta$ to $1/4$ in agreement
with \cite{kimKosterlitz1989,Kim91,L98}. This issue is important, because one
may speculate that discretized simulations cannot describe the local
singularities of continuum models, i.e. finite slopes may cause
corrections, responsible for the longstanding debate between field
theory and discrete model simulations.

In this paper we show that the converse is true. By performing very careful
corrections-to-scaling analysis on the model of Ref.~\cite{Kim91,kim2015}
we show that even in case of $N>1$ the rational numbers of \cite{kimKosterlitz1989,Kim91,L98}
can be excluded in the $L\to\infty$ limit.
Local slopes analysis shows, that the $N=1$
case has the smallest corrections and describes the \gls{kpz} universality scaling
the best. For $N>1$ corrections corresponding shorter wavelengths
are introduced. Our findings are in full agreement with the scaling
results obtained for ballistic growth models
\cite{AOF2014,Alves3d,Alvesunpub}.

\section{Models and simulation algorithms}

In order to enable long time surface growth simulations of large
systems, a multisurface-like parallel implementation of
the \gls{rsos} model has been created for \glspl{gpu}.
Two parallelization approaches have been combined as follows:

Since \glspl{gpu} feature a number of vector processors, multiples of $128$
realizations of the model were simulated simultaneously. This creates a
data-parallel workload, which can straightforwardly be vectorized. Each
\gls{simt} unit of the \gls{gpu} updates $128$ realizations, in which the
sequence of randomly selected coordinates for update is the same. This
correlation was broken by updating only half of the selected lattice sites in
each attempt.
If more realizations were simulated, different sets of $128$ realizations
evolved completely independently.

In order to handle large systems effectively a \gls{dd} was also
used to distribute the work of realizations among multiple \gls{simt} elements.
A double-tiling scheme was applied by splitting up the simulation cells into
tiles, split further into two sub-tiles along each spatial
direction~\cite{KONSH2012}. In the present two-dimensional problem this yields
$2^d=4$ sets of sub-tiles, each of which can be updated by multiple independent workers.
After each lattice sweep the origin of the \gls{dd} was moved randomly to eliminate
correlations.
Implementation details will be published elsewhere~\cite{msRsosUnpub}.

Roughening of $(2+1)$-dimensional \gls{rsos} surfaces was studied for restriction
parameters $N=1,3,5,7$, by starting from flat initial conditions. To obtain
estimates for the exponent $\beta$, the growth of surfaces was followed
up to $t=10^5$~\gls{mcs}, which is well before the correlation length approaches
the system sizes: $L=4096$, $8192$ and $9605$ studied here
(throughout this paper the time is measured in \gls{mcs}). The largest system size was
bounded by memory constraints, filling up \SI{12}{GB} of the NVIDIA K40 GPU, and
leaving some memory for the \gls{rng} states. The results were averaged over
$n=768, 128$ and $128$ realizations, respectively, where the latter two correspond
to only one multisurface run.

The exponent $\alpha$ was determined by a finite-size scaling analysis
of the saturation roughness of system sizes between $L=64$ and $L=512$.
To keep the noise amplitude constant we used domain sizes of $8\times8$ lattice sites.
We determined the interface width by averaging over $W(L,t)$ for times
$t\geq t_\mathrm{start}$ and for all samples. We checked whether the
averaged values belong to the steady state: $t > t_{\mathrm{steady}^*}$
by varying $t_\mathrm{start}$, the onset times of the measurements.
We estimated $t_{\mathrm{steady}^*}$ via the relation
\begin{equation}
 \label{eq:tSteadyApp}
 a_N\cdot L^\alpha = b_N\cdot t_{\mathrm{steady}^*}^\beta \ ,
\end{equation}
using the parameters $a_N$ and $b_N$, deduced from fitting in small systems.

In order to estimate the asymptotic values of $\alpha$ and $\beta$ for
$L\to\infty$ and $t\to\infty$, respectively, a local slopes analysis
of the scaling laws was performed~\cite{odorbook}. We calculated
the effective exponents
\begin{align}  \label{eq:aeff}
 \alpha_\mathrm{eff}\left(\frac{L - L/2}2\right) &= \frac{\ln W(L, t\to\infty) - \ln
W(L/2,t\to\infty)}{\ln(L)-\ln(L/2)} \\
\label{eq:beff}
\beta_\mathrm{eff}\left(\frac{t_i - t_{i/2}}2\right) &= \frac{\ln W(L\to\infty,t_i) - \ln
W(L\to\infty,t_{i/2})}{\ln(t_i) - \ln(t_{i/2})}\,.
\end{align}
In our studies the simulation time between two measurements is
increased exponentially
\begin{equation}
 t_{i+1} = (t_i+10)\mathrm{e}^m\quad,
\end{equation}
using $m=\num{.01}$ and $t_0 = 0$, while statistical uncertainties
are provided as $1\sigma$--standard errors,
defined as $\Delta_{1\sigma} x = \sqrt{\langle x^2\rangle - \langle
x\rangle^2}/(N-1)$.

\section{Surface growth results}

\subsection{The growth regime}

\begin{figure}
 \centering
 \includegraphics{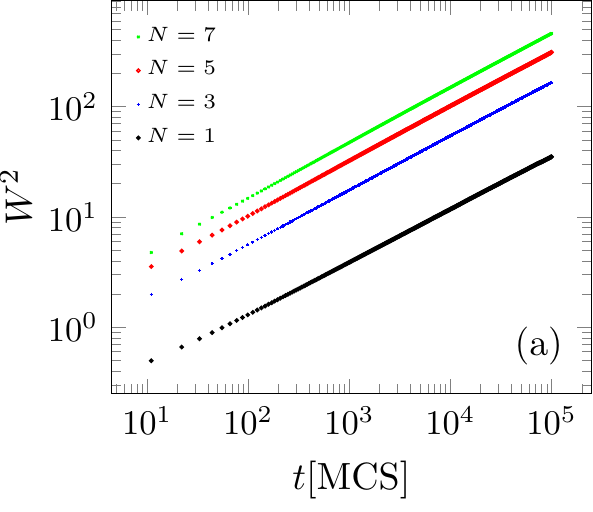}
 \includegraphics{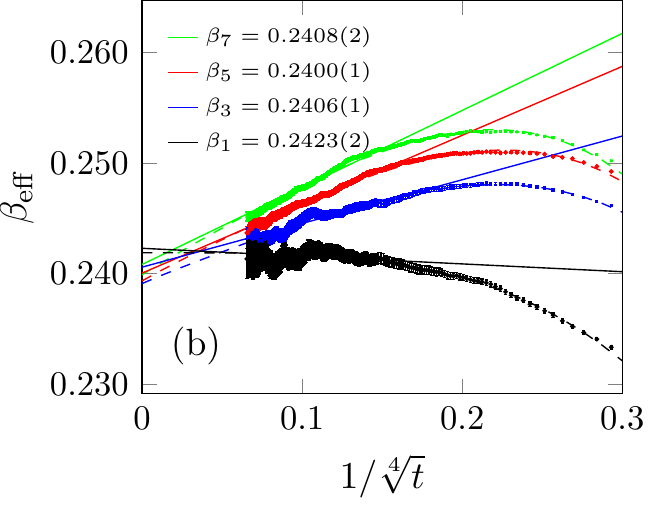}%
 \caption
 {\label{fig:rsosScaling}(Color online)
  (a) Squared roughness ($W^2$) of surfaces of size $V=4096^2$
  (256 realizations) in
  the scaling regime (error-bars are smaller than symbols).
  (b) Local slopes analysis of roughness scaling for
  size $V=8192^2$ (128 realizations). Straight lines are linear fits to the
  tail ($t\geq\SI{1260}{MCS}$),
  extrapolating to $t\to\infty$, assuming $\sqrt[4]t$ corrections.
  Uncertainties given for $\beta_N$ are errors of the singular linear fits
  displayed in the plot.
  The black dashed line is the \gls{pl} extrapolation for $N=1$.
  The dashed lines corresponding in color to the respective plots for $N>1$ are
  fits of the form~\eqref{eq:rsosBetaEffTwoPL}.  All \gls{pl} fits were
  performed for~$t\geq\SI{148}{MCS}$.
  Both figures show $N=1,3,5,7$ (bottom to top).
 }
\end{figure}

The growth of the surface roughness follows apparently the same, clear, \gls{pl}
for all considered $N$ (Fig.~\ref{fig:rsosScaling}(a)). The local slopes
plots (Fig.~\ref{fig:rsosScaling}(b)), using (\ref{eq:beff}),
show an effective growth exponent $\beta_\mathrm{eff}\approx\num{.25}$ for $N=5,7$ for
$t\le\SI{1000}{MCS}$ ($t^{-1/4}\approx\num{0.18}$), in agreement
with Kim's results~\cite{kim2015}. Later, the effective growth
exponent decreases for all $N>1$, followed over two
orders of magnitude in time in Fig.~\ref{fig:rsosScaling}.

Expecting independence of $\beta$ from $N$, it follows that the
asymptotic estimates $\beta_N$ should be the same. By assuming
\gls{pl} corrections to the asymptotic scaling
$W(L\to\infty,t) \propto t^\beta (1 + t^{-x})$,
we obtained a minimal variance of the $\beta_{N>1}$ estimates in case of
$x \simeq 0.25$. Therefore, we plotted our $\beta_\mathrm{eff}$ results
on the $\sim1/\sqrt[4]t$ scales, which makes the tails of the curves
straight in the $N\to\infty$ limit. Logarithmic corrections to scaling
were also tested, but they did not improve the extrapolations.

Table~\ref{tab:rsosScaling} lists the obtained
estimates for $\beta$ for the considered system sizes. Results for different $N>1$
are practically identical and are thus averaged to give a common value.
The case $N=1$ is listed separately, due to the different corrections
to scaling. For $N=1$, $\beta_\mathrm{eff}$ can be best extrapolated
by a \gls{pl} fit with $x=\num{0.90}(2)$.
This is in good agreement with the results of \cite{OAF2013},
where $x \simeq 0.96 \simeq 4\beta$ is reported, based on the
\gls{kpz} ansatz hypothesis.
This motivated us testing more general scaling forms,
with correction exponents multiple of $x=\beta$.
When we combined the effective exponent forms of $N=1$ and $N>1$,
\begin{equation}
 \label{eq:rsosBetaEffTwoPL}
 \beta_\mathrm{eff}(1/t) = \beta + a_1 / t^{4\beta} + a_2 / t^{\beta}\quad,
\end{equation}
with free parameters  $a_i$, fitting for $t\geq\SI{148}{MCS}$
resulted in good agreement for most of the growth region.
This is shown for $L=8192$ by the dashed lines in Fig.~\ref{fig:rsosScaling}(b).
From these extrapolations we obtained the estimates: $\beta_{N>1}=\num{0.2395}(5)$
and $\beta_{1}=\num{0.2415}(5)$.

As we can observe in Fig.~\ref{fig:rsosScaling}, the effective exponents
suffer from stronger corrections for $N>1$, than in the $N=1$ case.
Furthermore, our data suggest a possible oscillating convergence of 
$\beta_\mathrm{eff}$ for $N>1$, as reported in simulations of the 
\gls{bd} ~\cite{AOF2014}.
Extrapolations based on the form~\eqref{eq:rsosBetaEffTwoPL}, while
in good agreement within the observed region, are prone to over-fitting, 
where they can not cover all possible corrections. 
The values for $\beta_{N>1}$ are thus underestimated, if the effective 
exponents do indeed show oscillating convergence.

The estimates show no clear dependence on system size, thus it can
be safely assumed that all simulations are well within the scaling regime 
and do not suffer from finite-size effects. 
All results are within the margin of
error of the octahedron model $\beta=\num{.2415}(15)$~\cite{PhysRevE.84.061150}.
Most notably this is also the case for the estimates for $N>1$.
Statistical error measures for single extrapolations do not account for
systematic contributions such as from the choice of the extrapolation form or
the interval used for a fit. This can be clearly seen by the fact, that many
extrapolated values listed above do not agree with each other within such a
margin. The spread of these different estimates itself provides a more useful
estimate of the margin of error.
Overall, the presented data support $\beta=\num{.241}(1)$.

Since the curves in Fig.~\ref{fig:rsosScaling} correspond to the same $L$
and sample size $n$, one can observe that the \gls{snr}, the ratio between
the interface width and the sample variance, increases with $N$.
For $N=7$ this is higher by a factor of $\sim\num{3.6}$, while for $N=3$ the
\gls{snr} is about $\sim\num{2.5}$ bigger than that of the $N=1$ result.
Presumably, the decrease of relative noise level is the consequence
of a kind of self-averaging, since systems with larger allowed $N$
accommodate more surface information than smaller ones.
It is tempting to exploit this property by choosing larger height differences
in the simulations, even if this can be implemented less efficiently.

\begin{table}
 \caption{\label{tab:rsosScaling}
  Extrapolated $\beta$ results for different $N$. For $N=1$ figures in the parentheses are
  fit errors from \gls{pl} extrapolations. For $N>1$, given margins are $1\sigma$
  standard errors from averaging over $N=3,5$ and $7$.
 }
 \centering
 \begin{tabular}{lrrr}
  \hline
  $L$             & 4096 & 8192 & 9605 \\
  \hline
  $\beta_1$       & 0.2412(1) & 0.2418(1) & 0.2415(1) \\
  $\beta_{N>1}$ & 0.2404(3) & 0.2405(3) & 0.2410(3) \\
  \hline
 \end{tabular}
\end{table}

\subsection{The steady state} \label{sec:ss}

Direct fitting of the finite size scaling form
\begin{equation}
 W_\mathrm{sat}(L)\sim L^\alpha, \label{eq:rsosFiniteSizeScaling}
\end{equation}
for $32 \leq L \leq 512$ and $t_\mathrm{start}=50t_{\mathrm{steady}^*}$
yields the following estimates
\[
 \alpha_\mathrm{fit} = \left\{\begin{tabular}{rrr}
  0.392(1) & \num{.392}(5) & N=1 \\
  0.401(2) & \num{.400}(4) & N=3 \\
  0.402(2) & \num{.401}(4) & N=5 \\
  0.402(2) & & N=7 \\
 \end{tabular}
 \right.
\]
For comparison, Kim's results~\cite{kim2015} are shown in the second column.
When we decrease $t_\mathrm{start}$ our values decrease slightly but
fall inside the error margins if $t_\mathrm{start} \geq 2t_{\mathrm{steady}^*}$.
So, direct fits match perfectly those of
~\cite{kim2015}, obtained by sequential Monte Carlo updates.

\begin{figure}
 \centering
 \includegraphics{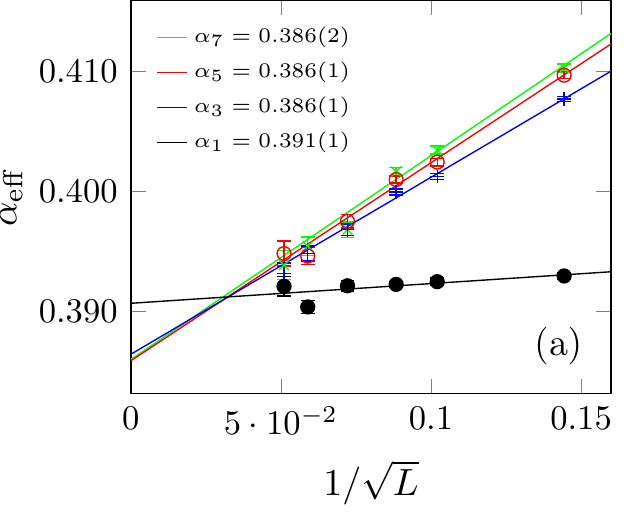}
 \includegraphics{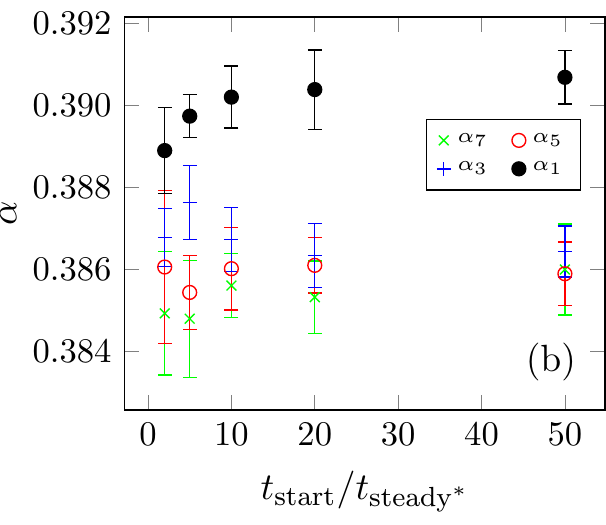}
 \caption{\label{fig:rsosSteadyEE}(Color online)
  (a) Local slopes of finite-size scaling analysis with
  $N=1,3,5,7$. Error bars are propagated $1\sigma$ errors. Straight lines are
  linear fits to extrapolate to infinity, uncertainties given for $\alpha_N$ are
  pure fit errors.  Steady-state data taken for
  $t>t_\mathrm{start}=50t_{\mathrm{steady}^*}$ (see text).
  (b) Dependence of extrapolated $\alpha$ on $t_\mathrm{start}$ is weak.
  Both figures: Sample sizes are at least 1024--2048 realizations and $\geq8192$
  realizations for $L\leq64$. All system sizes taken into account for
  finite-size scaling are listed in Fig.~\ref{fig:rsosSteadyCollL}, where
  the considered time scales can also be read off.
 }
\end{figure}

However, if the $L=32$ data are excluded, our estimates become
significantly lower, warning for strong corrections to scaling.
This can also be seen with the help of the effective exponents
in Fig.~\ref{fig:rsosSteadyEE} calculated by (\ref{eq:aeff}).
There is a clear tendency for $\alpha_\mathrm{eff}$ to decrease
as we increase the system size for the $N>1$ cases.
The approach to $L\to\infty$ is nonlinear, but the number of points
is insufficient for \gls{pl} extrapolations to produce consistent estimates.
We plotted the $\alpha_\mathrm{eff}(L)$ results on the $1/\sqrt{L}$ scale,
resulting in points that can be settled on straight lines. Linear
extrapolation to asymptotically large sizes yields:
\[
 \alpha = \begin{cases}
  0.391(1) & N=1 \\
  0.386(1) & N>1 \\
 \end{cases}
\]
Corrections to finite-size scaling (\ref{eq:rsosFiniteSizeScaling}) in case of $N=1$ are small,
explaining the good agreement between local slopes analysis and the direct fit.
The slight difference between the results for $N=1$ and $N>1$ may be attributed
to the fact that our data points are not from deep enough in the steady state.
This might also explain the disagreement with the results of a recent
study~\cite{PaganiParisi2015}, which reported $\alpha=\num{.3869}(4)$ for $N=1$.
There is a further uncertainty of the extrapolation to $L\to\infty$,
which is not accounted for by the fit errors.
With the assumption of an intrinsic width: $W^2_i = 0.2$~\cite{Alves3d}, the
local slopes analysis shows stronger corrections to scaling, therefore we did
not apply this in our study.

The observation of stronger corrections for larger $N$s
is consistent with a recent analysis of the \gls{bd}.~\cite{AOF2014}
This study found that corrections to scaling, for both $\alpha$ and $\beta$, are
reduced, when the \gls{bd} surface is smoothened by binning of the surface
positions before analysis, thereby decreasing the height differences between
neighboring sites. Binning of the surface did not change the universal
behavior; it only decreased non-universal corrections. The corrections
produced even an oscillatory approach to the asymptotic values of the
exponents. This can explain why our simple extrapolations of $\alpha_\mathrm{eff}$
(Fig.~\ref{fig:rsosSteadyEE}) and $\beta_\mathrm{eff}$ (Fig.~\ref{fig:rsosScaling})
for $N>1$ undershoot those of $N=1$.

All of our estimates up to $N\leq7$, obtained by the local slopes analysis,
are in the range $\alpha=\kpzRsosAlpha$, which clearly excludes $\alpha = 2/5$.
Plugging our $\alpha$ and $\beta$ results into the scaling relation~\eqref{zlaw}
we get the dynamical exponent estimates $z_{N=1}=\num{1.61}(2)$ and $z_{N>1}=\num{1.60}(2)$,
respectively. The scaling law following from the Galilean invariance is satisfied
with these exponents both for $N=1$: $\alpha + z = \num{2.01}(2)$
and $N>1$: $\alpha + z = \num{1.99}(2)$ within error margins.

\begin{figure}
 \centering
 \includegraphics{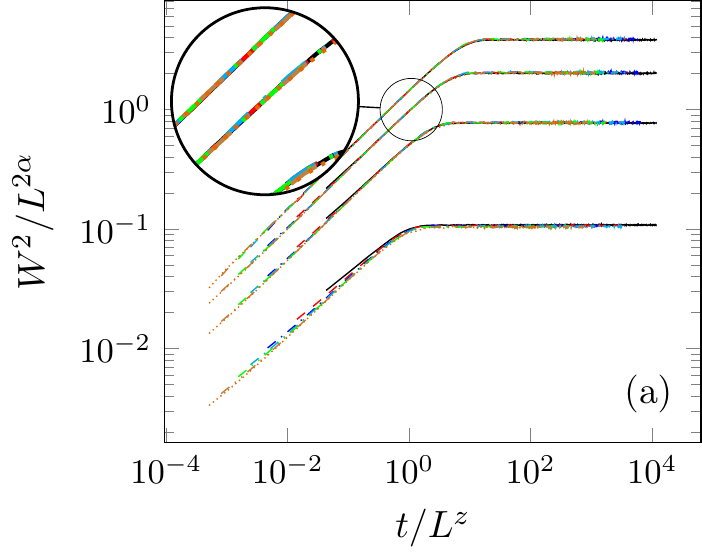}
 \includegraphics{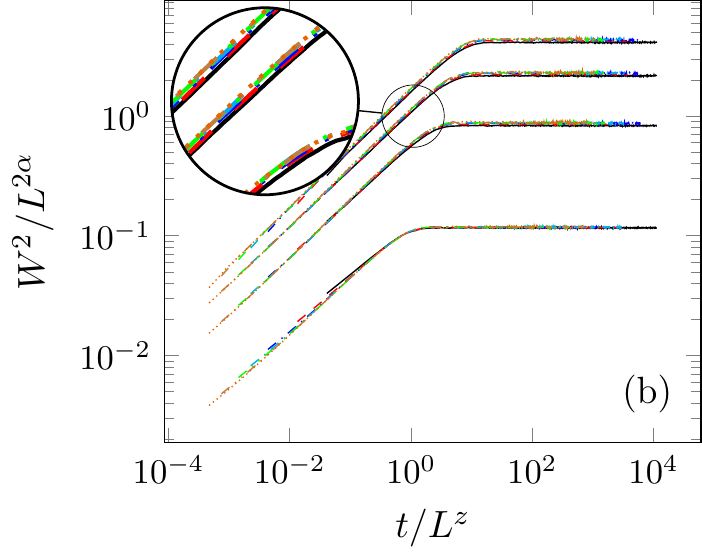}
 \caption{\label{fig:rsosSteadyCollL}(Color online)
  Collapse of squared roughness in the steady state for $N=1,3,5,7$
  (from bottom to top). Panel (a) shows a perfect collapse for $N>1$,
  using $\alpha=\num{.4}$ and $\beta=\num{.25}$ ($z=\alpha/\beta=\num{1.6}$).
  Panel (b) shows a collapse using $\alpha=\num{.389}$ and
  $\beta=\num{.241}$ ($z\approx\num{1.61}$). This looks perfect for
  $N=1$, but not for $N>1$.
 }
\end{figure}

We have also tested the scaling form~\eqref{FV-forf} numerically by using our
$\alpha$ and $\beta$ values. As Fig.~\ref{fig:rsosSteadyCollL} shows,
good data collapses can be obtained for $N>1$ and even a perfectly
looking one for $N=1$. For $N>1$ in the growth regime a perfect one
can also be achieved assuming the values suggested by Kim and Kosterlitz~\cite{kimKosterlitz1989}
(Fig.~\ref{fig:rsosSteadyCollL}(a)). This can be understood
by taking into account the corrections to scaling we explored above.
Effective exponents for early times and small systems agree with the
conjecture by~\cite{kimKosterlitz1989} and indeed the most strongly
outlying curves in Fig.~\ref{fig:rsosSteadyCollL}(a), correspond
to smaller systems.

Moments of the width and
height distributions are defined as:
\begin{align}
 \Phi^n_L[\varphi_L] &= \int\limits_0^\infty \left(\varphi_L -\langle
\varphi_L\rangle\right)^n P_L(\varphi_L)\,\mathrm{d}\varphi_L\quad,
\intertext{where $P_L( \varphi_L)$ denotes the probability distribution corresponding
 to the interface observable $\varphi_L$. We calculated some
standard measures of the shape, the skewness}
S_L[\varphi_L]&= \langle \Phi^3_L[\varphi_L]\rangle / \langle \Phi^2_L[\varphi_L]\rangle^{3/2}
\intertext{and the kurtosis}
Q_L[\varphi_L]&= \langle \Phi^4_L[\varphi_L]\rangle / \langle
\Phi^2_L[\varphi_L]\rangle^{2} - 3\quad,
\end{align}
in the steady state. These measures were shown to be universal in \gls{kpz}
models~\cite{MPPR02,FORWZ1994}. 

\begin{figure}
 \centering
 \includegraphics{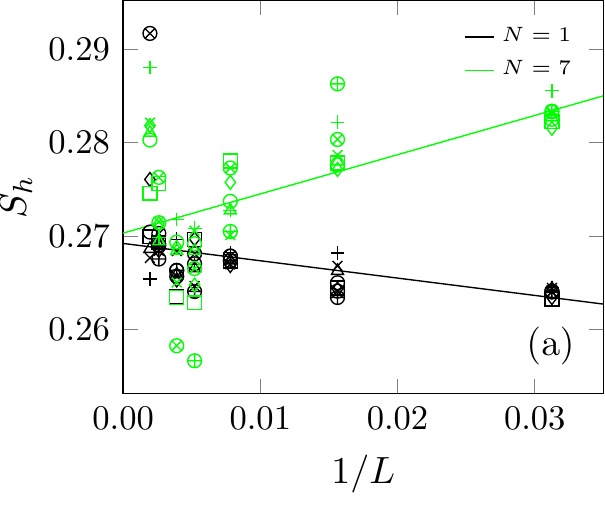}
 \includegraphics{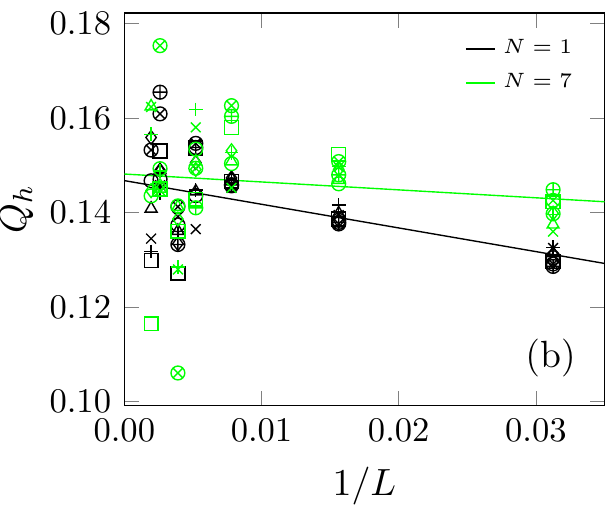}
 \caption{\label{fig:rsosSteadySkH}(Color online)
  Skewness $S_h$ (a) and kurtosis $Q_h$ (b) of the height distribution in
  the steady state plotted over the inverse lateral system size. Values are
  plotted only for $N=1$ (black) and $N=7$ (green) for the sake of clarity. The
  straight lines are linear fits, included to guide the eye. Different symbols
  indicate different ratios $t_\mathrm{start}/t_{\mathrm{steady}*} \geq 2$. A
  key is not provided for the symbols, because there is no correlation with this
  parameter.
 }
\end{figure}

The obtained values for the width-distribution $P_L(W^2(L))$ show no
significant dependence on $N$ nor $L$, our
best results are $S=1.70(1)$ and $Q=5.38(4)$, in good agreement
with those of~\cite{FDAAR05}.

For the distribution of surface heights, a weak correlation with the system size
can be observed in Fig.~\ref{fig:rsosSteadySkH}. Heights were averaged in the
steady state starting at different times $t_\mathrm{start} >
t_{\mathrm{steady}*}$ (indicated by different symbols in the figure), but no
dependence can be observed. Our results $S_h=\num{0.270}(5)$ and
$Q_h=\num{0.15}(1)$ are in agreement with the ranges given
in~\cite{PaivaReis2007} and especially with the values $S_h=\num{0.26}(1)$ and
$Q_h=\num{0.134}(15)$ reported in references~\cite{MPP2000,AaroReis2004}. Thus
all cumlant values are within error margins of the KPZ universality class
irrespectively of $N$.

\subsection{Consistency of fine-size scaling with respect to \gls{dd}}

Since we used a parallel DD in our simulations we have also checked for
dependence of the results on the applied scheme. We performed
additional finite-size scaling studies with domains of $16\times16$
and $6(+1)\times10(+1)$ lattice sites.
The figures in the parentheses refer to irregular tiling of the system.
This is the consequence of the fact, that lattices cannot be divided into
domains with a lateral size of six (or ten) sites without remainder,
thus a subset of domains have larger lateral size to compensate it.
This configuration results from dividing the system into multiples of
$5\times3$ tiles, in order to achieve optimal load balancing on NVIDIA GTX
Titan Black \glspl{gpu}. In both cases the smallest considered system size
was $L=64$ to avoid unreasonable \gls{dd}. Another test was done using
$3(+1)\times5(+1)$ sized domains. These tiles turned out to be too small
to give correct results, expressed by failing data collapses,
thus we do not consider them in the following discussion.

\begin{figure}
 \centering
 \includegraphics{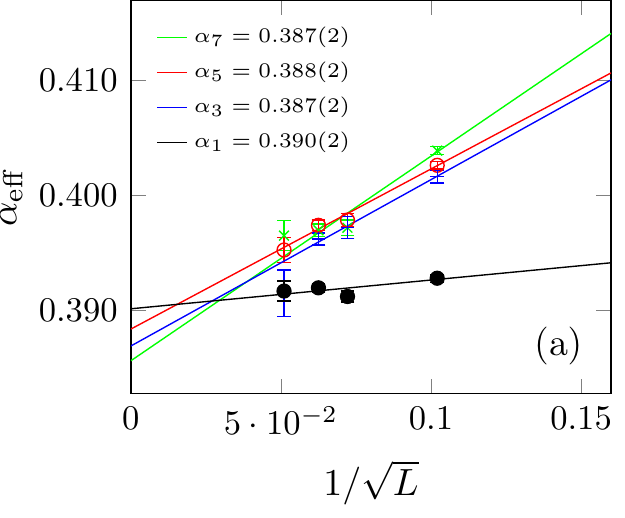}
 \includegraphics{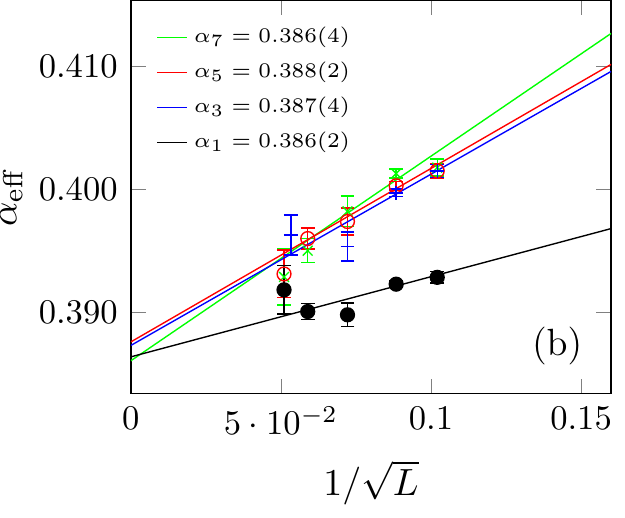}
 \caption{\label{fig:rsosSteadyDDEE}(Color online)
  Local slopes of finite-size scaling analysis for
  $N=1,3,5,7$. Error bars are propagated $1\sigma$ errors. Straight lines are
  linear fits to extrapolate to infinity, uncertainties given for $\alpha_N$ are
  pure fit errors.  Steady-state data are taken for
  $t>t_\mathrm{start}=50t_{\mathrm{steady}^*}$ (see text).
  (a) \gls{dd} domains containing $6(+1)\times10(+1)$ sites. Sample sizes are
  at least $512$ realizations, for $N=5,7$ and sizes $L=64$ and $128$, $n=16384$
  $n=8192$ are used.
  (b) \gls{dd} domains containing $16\times16$ sites. For $L=512$ the sample
  contains 256 realizations, for other system sizes at least 512 samples are
  included.
 }
\end{figure}

The differences among the results of the considered \gls{dd} configurations
were significant neither in the data collapses nor in the finite-size scaling fits.
The most sensitive quantity proved to be the effective roughness exponent,
shown in Fig.~\ref{fig:rsosSteadyDDEE}.
Sample sizes of this test were smaller than those of Sec.~\ref{sec:ss},
making the extrapolations less reliable. Still, all
estimates derived from this data are consistent with the estimate
$\alpha=\kpzRsosAlpha$. Even the results of irregular, non-square
\glspl{dd} do not deviate significantly,
although small systematic errors might be present.

\section{Conclusions}

Extensive numerical simulations have been performed for
$(2+1)$-dimensional \gls{rsos} models with variable height
difference restrictions. Careful correction to
scaling analysis has provided numerical evidence that the universal
surface growth exponents agree with the most precise values known for the
$(2+1)$-dimensional \gls{kpz} class.
These estimates, $\alpha=\num{0.390}(4)$ and $\beta=\num{0.241}(1)$, exclude the
rational values $\alpha=4/10$ and  $\beta=1/4$, conjectured
by~\cite{kimKosterlitz1989,Kim91,L98,kim2015}. Our results
support the generalized \gls{kpz} ansatz, which takes finite-time
corrections into account and predicts exponents $x$ that are
multiples of $\beta$ \cite{OAF2013}. We found
$x = 0.90(2)$ for $N=1$ and $x\simeq 0.25$ for $N>1$. 

We have shown that by increasing the local height differences we obtain
better \gls{snr} in the simulations, but stronger corrections to scaling, 
which can confuse numerical analysis based on simple \gls{pl} fitting. 
Therefore, smaller step-sized models, like the octahedron model~\cite{PhysRevE.84.061150} 
describe better the asymptotic, long-wave-scaling behavior of the 
\gls{kpz} universality class.
Our conclusions for scaling corrections are in agreement with those obtained
for ballistic growth models~\cite{AOF2014,Alves3d}. According to our knowledge
oscillating convergence of effective exponents has not yet been observed in \gls{rsos}
models, necessitating further investigations. We also provided estimates
for the skewness $S=\num{1.70}(1)$ and the kurtosis $Q=\num{5.38}(4)$ of the 
surface width distributions as well as $S_h=\num{0.270}(5)$ and
$Q_h=\num{0.15}(1)$ for the height distributions, both in the steady state.
Our simulations have been performed
using multisurface \gls{gpu} \gls{simt} algorithms with origin moving domain
decomposition.
The results have been justified by varying the tile sizes. A sustained
performance of~$\simeq\num{1.1e10}$ deposition attempts per second could be
achieved running on a single NIVIDIA GTX Titan Black \gls{gpu}.  This opens up
the possibility for precise \gls{rsos} simulations in higher dimensions.

\vskip 1.0cm

\noindent
{\bf Acknowledgments:}\\

We thank S.~Alves for sending us the correction-to-scaling plot of exponent 
$\alpha$ of the three-dimensional ballistic growth and S.~C.~Ferreira and T.~Halpin-Healy
for useful comments.
Support from the Hungarian research fund OTKA (Grant No.~K109577), the
Initiative and Networking Fund of the Helmholtz Association via the W2/W3
Programm \mbox{(W2/W3-026)} and the International Helmholtz Research School
NanoNet \mbox{(VH-KO-606)} is acknowledged.
We gratefully acknowledge computational
resources provided by the HZDR computing center, NIIF Hungary and the Center for
Information Services and High Performance Computing (ZIH) at TU Dresden.
We acknowledge support by the GCoE Dresden.
J.~K.~thanks M.~Weigel from
Coventry University for providing a guest position, co-funded through the
Erasmus+ program via the Leonardo-B\"uro Sachsen.

 \bibliography{bib}

\end{document}